\documentclass[aps, twocolumn,floats, showpacs]{revtex4-1}
\usepackage{amsmath,amssymb,graphicx}
\usepackage{epsfig}
\usepackage{graphicx}
\usepackage{enumerate}
\usepackage{natbib}

\newcommand{\beq}{\begin{equation}}
\newcommand{\eeq}{\end{equation}}

\newcommand{\bea}{\begin{eqnarray}}
\newcommand{\eea}{\end{eqnarray}}

\begin{document}



\title
{Qubit-mediated energy transfer between thermal reservoirs: beyond Markovian Master equation}
%
\author{Dvira Segal}
\affiliation{Chemical Physics Theory Group, Department of Chemistry,
University of Toronto, 80 Saint George St. Toronto, Ontario, Canada
M5S 3H6}

\date{\today}
\begin{abstract}
We study qubit-mediated energy transfer between two electron reservoirs
by adopting a numerically-exact influence functional path-integral method.
This non-perturbative technique allows us to study the system's dynamics
beyond the weak coupling limit.
Our simulations for the energy current indicate 
that  perturbative-Markovian Master equation predictions
significantly deviate from exact numerical results already at
intermediate coupling, $\pi \rho \alpha_{j,j'}\gtrsim 0.4$, where
$\rho$ is the metal (Fermi sea) density of states, taken as a
constant, and $\alpha_{j,j'}$ is the scattering potential energy of
electrons, between the $j$ and $j'$ states. Markovian Master
equation techniques should be therefore used with caution beyond the
strictly weak subsystem-bath coupling limit, especially when a
quantitative knowledge of transport characteristics is desired.
\end{abstract}


\maketitle


\section{Introduction}
\label{Sec1}

Quantum impurity models, including a subsystem interacting with a reservoir,
were proven useful in describing and predicting many physical phenomena.
The spin-boson model \cite{SB}, representing the dynamics of a
single charge on two states coupled to a dissipative bath, e.g., a
solvent, exhibits rich phenomenology, including various phase
transitions \cite{Weiss}. It is relevant for modeling charge
transfer reactions in biological systems \cite{Weiss},
photosynthesis \cite{bioSB}, the Kondo problem for magnetic
impurities \cite{Kondo}, and quantum information processing in
superconducting Josephson tunneling junction qubits \cite{Shnirman}
or nitrogen-vacancy centers in diamonds \cite{NV}. A variant of the
spin-boson model is the  spin-fermion model, where a qubit, referred
to as a spin or a two-level system, interacts with a metallic-fermionic environment. 
This model is also related to the Kondo model \cite{Kondo}, only
lacking direct coupling of the reservoir degrees of freedom to
spin-flip processes. The generalization of the equilibrium
spin-fermion model, to include more than one Fermi bath, provides a
minimal setting for the study of dissipation and decoherence effects
under the influence of an out-of-equilibrium environment
\cite{MitraSpin1, MitraSpin2, SMarcus,SF}.

\begin{figure}[htbp]
\vspace{-2mm}
{\hbox{\epsfxsize=160mm \hspace{8mm}\epsffile{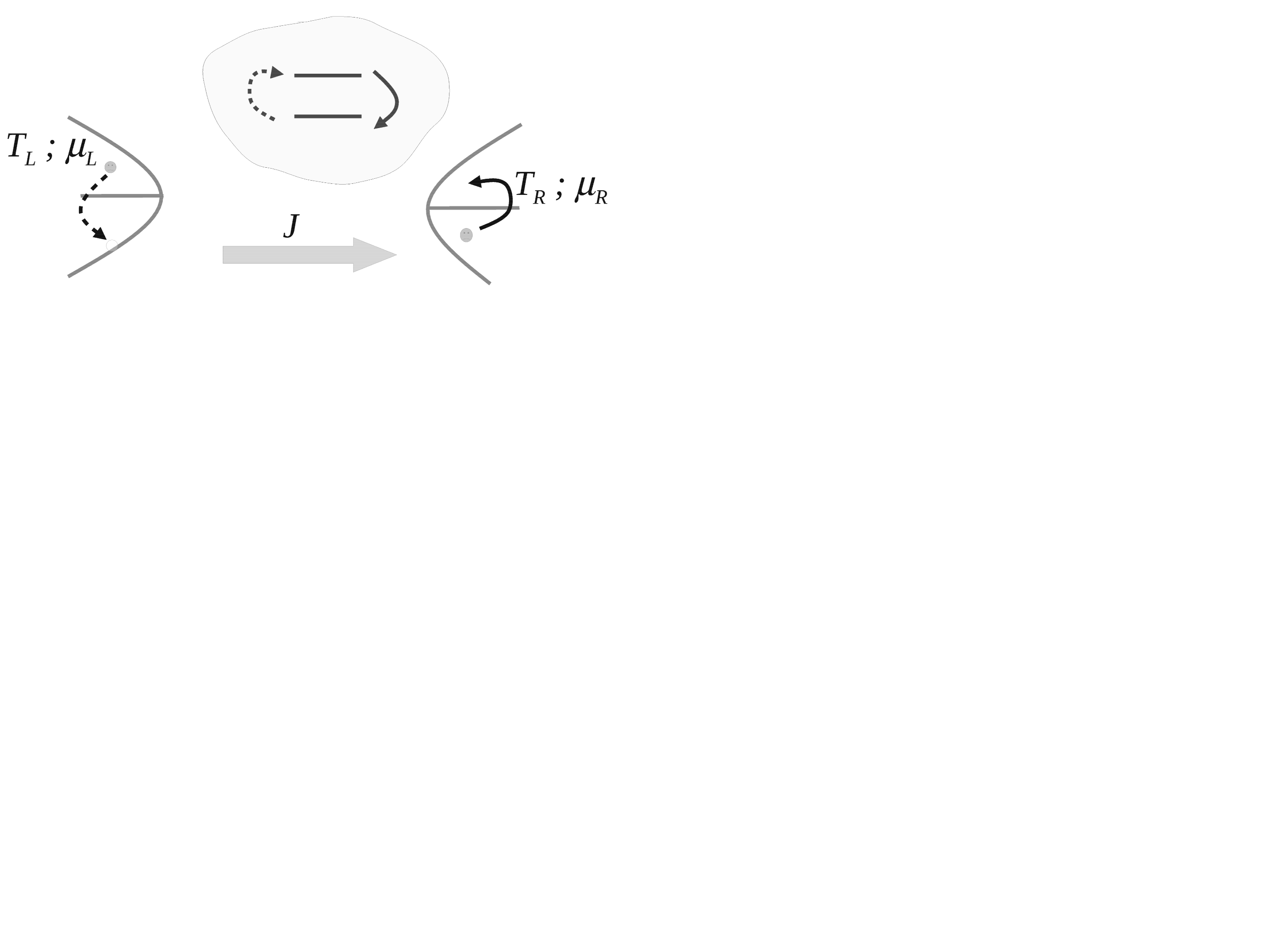}}}
\caption{A schematic representation of our model system. Electron
transfer between the metals is blocked, but energy current is
flowing through an excitation/de-excitation of the intermediate
anharmonic (two-state) quantum system. The curved arrows represent
energy transfer processes between the leads and the intermediating
subsystem. In our work here we set $\mu_L=\mu_R$ and take
$T_L>T_R$.}
 \label{FigS}
\end{figure}

In this work, we use the two-bath spin-fermion model and investigate
energy exchange between two metals, mediated by the
excitation/relaxation of a  nonlinear quantum system, a qubit. For a
scheme of this setup, see Fig. \ref{FigS}. Physically, our model can
describe the process of radiative heat transfer between metals
\cite{Pekola1,Pekola2,radiation}, and it can be realized within a
superconducting Josephson junction circuit \cite{heat1,SF,KochSC}.
We simulate  the energy current characteristics of the
nonequilibrium spin-fermion model in a large parameter range of
coupling strengths by means of an influence-functional path-integral
(INFPI) technique developed in Refs. \cite{IF1,IF2}. This
numerically-exact method is built about the basic observation, that
in out-of-equilibrium (and/or finite temperature) situations bath
correlations have a finite range, allowing for their truncation
beyond a memory time dictated by the voltage-bias and the
temperature \cite{MitraSpin2, SMarcus, QUAPIRev}. Taking advantage
of this fact, an iterative-deterministic time-evolution scheme can
be developed, where convergence with respect to the memory length
can in principle be reached.


Our main objective here is to explore the qubit-mediated energy
current characteristics beyond standard perturbative methods.
Particularly, we would like to find when the Golden-Rule-type
Markovian Master equation method provides a correct (quantitative or
qualitative) description of the exact behavior. This task is
important since Master equation tools have been extensively adopted
for studying problems in charge, spin, and energy transfer
phenomenology in quantum dots and molecular junctions, see for
example \cite{BrandesRev, MitraMaster,MasterWu,radiation, Brandes1,
Koch1,*Koch2,*Koch3,*Koch4,Teemu,
Teemuheat,OjanenEff,Gernot,Motonen}.

The plan of the paper is as follows. In Sec. II we present the
nonequilibrium spin-fermion model. We provide expressions for
observables of interest in Sec. III. The two methods confronted,
INFPI and Markovian Master equation, are discussed in Sec. IV, with
results included in Sec. V. Sec VI. concludes. For simplicity, we
use the conventions $\hbar\equiv 1$, electron charge $e\equiv 1$,
and Boltzmann constant $k_B=1$.


\section{Model}
\label{SecModel}

The system of interest comprises of two metallic leads, $\nu=L,R$, prepared at
different temperatures but at the same chemical potential. These metals
are connected indirectly, by a nonlinear quantum unit, a
two-level subsystem.
The Hamiltonian includes three contributions,
\bea
H=H_S+H_F+V,
\eea
where
\bea
H_S&=&\Delta \sigma_z,
\nonumber\\
H_F&=&H_L+H_R, \,\,\,\,\,\, H_{\nu}=\sum_{j\in \nu}\epsilon_j
c_{\nu,j}^{\dagger} c_{\nu,j}
\nonumber\\
V&=&V_L+V_R, \,\,\,\,\,\,
V_{\nu}=\sigma_x\sum_{j,j'}\alpha_{\nu,j; \nu, j'}c_{\nu, j}^{\dagger}c_{\nu, j'}.
\label{eq:H}
\eea
The subsystem $H_S$ includes two states, $|0\rangle$ and
$|1\rangle$, with a tunneling splitting  $2\Delta$. It can be
realized within a nonlinear resonator mode, or it can represent an
impurity in a solid-state environment. This subsystem interacts with
two fermionic reservoirs ($H_F$) where $c_{\nu,j}^{\dagger}$
($c_{\nu,j}$) creates (annihilates) an electron at the $\nu=L,R$
metal lead with momentum $j$, disregarding the electron spin degree
of freedom. The qubit-metal interaction term $V$ couples scattering
events within each metal to transitions within the subsystem. For
simplicity, we assume that the coupling constants $\alpha_{\nu}$ are
energy independent and real numbers. Note that we do not allow for
charge transfer processes between the two metals, assuming the
tunneling barrier is  high. However, energy is transferred between
the two metals, mediated by the excitation of the intermediate
nonlinear quantum system, see  Fig. \ref{FigS}. Using the
Hamiltonian form (\ref{eq:H}), the subsystem dynamics and the energy
current can be readily attained within a Markovian Master equation,
as we explain in Sec. IV.B

The Hamiltonian (\ref{eq:H}) can be transformed into the standard spin-fermion model
of zero energy spacing with a unitary transformation,
\bea
U^{\dagger}\sigma_z U=\sigma_x, \,\,\,\,\,  U^{\dagger}\sigma_x U=\sigma_z,
\eea
where $U=\frac{1}{\sqrt{2}}(\sigma_x+\sigma_z)$. The transformed
Hamiltonian $H_{SF}=U^{\dagger} HU$ includes a $\sigma_z$-type
electron-spin coupling,
\bea H_{SF}&=&\Delta \sigma_x+ \sum_{\nu,j}\epsilon_j
c_{\nu,j}^{\dagger} c_{\nu,j} \nonumber\\
&+&\sigma_z\sum_{\nu,j,j'}\alpha_{\nu,j; \nu, j'}c_{\nu,
j}^{\dagger}c_{\nu, j'}. \label{eq:HH} \eea
In this representation, the dynamics can be conveniently simulated with INFPI,
a brief discussion is included in Sec. IV.A.

\section{Observables}

We assume a factorized initial state with the total density matrix
$\rho(0)=\rho_S(0) \otimes \rho_F$. Here $\rho_S$ denotes the
reduced density matrix of the subsystem. The reservoirs' density matrix at time
$t=0$ is given by $\rho_F=\rho_L\otimes\rho_R$, and these
states are canonical,
$\rho_{\nu}=e^{-\beta_{\nu}(H_{\nu}-\mu_{\nu}N_{\nu})}/{\rm
Tr_{\nu}}[e^{-\beta_{\nu}(H_{\nu}-\mu_{\nu}N_{\nu})}]$. Here
we trace over the $\nu$ reservoirs degrees of freedom. In our
simulations below we take $\mu_L=\mu_R$, but assume different
initial temperatures, $T_L\neq T_R$. We refer to this setup as a
``nonequilibrium environment" since the two reservoirs are
prepared in different states.
%
At time $t=0$ we put into contact the two Fermi baths through the
quantum subsystem, and follow the evolution of the reduced density
matrix and the energy current, to steady-state.
Since the energy current in the system is driven by a temperature bias, we
can also refer to the energy current here as a ``heat current".

The time evolution of the reduced density matrix is obtained by
tracing $\rho$ over the fermionic reservoirs' degrees of freedom,
\bea
\rho_S(t) = {\rm Tr}_F \left[ e^{-iHt}\rho(0)e^{iHt}\right].
\label{eq:rhoS}
\eea
The definition of the energy current operator is more subtle
\cite{Wu,ThossCPL}, as different-plausible choices provide distinct
results in the short time limit. At long time, in the steady-state
limit, these definitions yield the same value. Here, we follow the
analysis of Ref. \cite{Wu}, and define the energy current operator,
e.g., at the left contact, as
\bea
\hat J_{\nu}=\frac{i}{2}[H_L-H_S,V_L].
\eea
In steady-state the expectation value of the interaction is zero,
and we reach the  relation,
\bea {\rm Tr}\left[\rho \frac{\partial V_L}{\partial t}\right]\equiv
\left \langle \frac{\partial V_L}{\partial t}\right \rangle= i
\langle [H_S+H_L,V_L] \rangle=0. \label{eq:App} \eea
Note that we have assumed that $[V_L,V_R]=0$. Using Eq.
(\ref{eq:App}) we reach the following expression for the averaged
energy current (valid in the long time limit),
\bea
\langle J_L\rangle \equiv {\rm Tr}_S{\rm Tr}_F[\hat J_L\rho(t)] =-i\langle [H_S,V_L]\rangle.
\label{eq:J2}
\eea
This commutator can be readily evaluated to yield,
\bea
[H_S,V_L]=2i\Delta\sigma_y\sum_{l,l'}\alpha_{L,l;L,l'}c_{L,l}^{\dagger}c_{L,l'},
\eea
leading to
\bea \langle J_L\rangle=
2\Delta {\rm Tr}_S\left[ \sigma_y{\rm Tr_F}[A_L\rho(t)]\right], \eea
with the bath operator $A_L\equiv \sum_{l,l'} \alpha_{L,l;L,l'}c_{L,l}^{\dagger}c_{L,l'}$.
We now define a subsystem operator
\bea A_S(t)\equiv{\rm Tr}_F[A_L\rho(t)]= {\rm
Tr}_F[e^{iHt}A_Le^{-iHt}\rho(0)], \label{eq:Ast} \eea
and express the current using its matrix elements
\bea \langle J_L\rangle=2\Delta [-i (A_S(t))_{1,0} + i
(A_S(t))_{0,1} ]. \label{eq:curr} \eea
We emphasize that this expression is designed to provide the steady-state
value and not the transients, given our assumption
(\ref{eq:App}).

The two operators, $\rho_S(t)$ and $A_S(t)$, are subsystem operators.
They are simulated in the next section directly, using INFPI, or studied in a perturbative manner,
under the Markovian limit,
to provide Kinetic-type expressions.


\section{Methods}


\subsection{Path-integral simulations}

The principles of the INFPI approach have been detailed in Refs.
\cite{IF1,IF2}, where it has been adopted for investigating
dissipation effects in the nonequilibrium spin-fermion model and
charge occupation dynamics in the interacting Anderson model. Other
applications include the study of the intrinsic coherence dynamics in a
double quantum dot Aharonov-Bohm interferometer \cite{AB},
exploration of relaxation and equilibration dynamics in finite metal
grains \cite{Kunal1,*Kunal2}, and the study of electron-phonon
effects in molecular rectifiers \cite{ETIF}.

Here, using INFPI, we can directly simulate both the dynamics of the
reduced density matrix $\rho_S(t)$ [Eq. (\ref{eq:rhoS})], and the
time evolution of subsystem expectation values, particularly
$A_S(t)$ [Eq. (\ref{eq:Ast})], which can be used to obtain the
energy current $\langle J_L\rangle$, Eq. (\ref{eq:curr}). In
practice, for achieving fast convergence, we have simulated directly
the averaged current
\bea
J=\frac{1}{2}(\langle J_L\rangle - \langle J_R\rangle).
\eea
The negative sign in front of $\langle J_R\rangle$ arises from our
sign convention; the current $\langle J_{\nu}\rangle$ is defined from
the $\nu$ reservoir, into the junction.

Algorithmic details of the INFPI method were recently elaborated in
Ref. \cite{ETIF}, thus we only include the main principles here. The
algorithm is based on a Trotter  breakup of a short-time time
evolution operator into two parts: a (simple) time evolution term
that depends on the subsystem Hamiltonian, and a term that
accommodates the reservoirs Hamiltonians, and their interactions
with the subsystem. Collecting the contribution of the latter terms
along the subsystem path, we construct the so called ``influence
functional" (IF), which involves nonlocal dynamical correlations.
The IF has an analytical form in some special cases \cite{QUAPIRev};
in the present model its form is only known in the weak-intermediate
coupling limit \cite{MitraSpin2,SMarcus}, thus we evaluate it
numerically by energy-discretizing the Fermi sea.

The main conceptual element behind the INFPI approach is the
observation that at finite temperatures and/or nonzero chemical
potential bias bath correlations exponentially decay in time,
allowing for their truncation beyond a memory time $\tau_c$. The
dynamics can then be achieved by defining an auxiliary density
matrix, or more generally, a subsystem operator [e.g., $A_S(t)$ of
Eq. (\ref{eq:Ast})],  on the time-window $\tau_c$. This nonlocal
object can be iteratively evolved from the subsystem-bath factorized
initial condition, to the present time $t$.

This path-integral method involves three numerical parameters:
(i) the number of states used in the discretization of
each Fermi sea $L$, (ii) the time step adopted in the Trotter breakup $\delta t$,
and (iii) the memory time accounted for, $\tau_c$, beyond which the IF,
accommodating the effect of the reservoirs on the
subsystem, is truncated. Convergence of INFPI is verified by
confirming that results are insensitive to the
reservoirs discretization, the finiteness of the time step, and the memory size
$\tau_c=N_s\delta t$, with $N_s$ as an integer.
It should be noted that minimizing the Trotter breakup error, taking
$\delta t\rightarrow 0$, conflicts with the need to cover the memory
time-window $\tau_c$, since the parameter $N_s$ has to be increased
until convergence is reached. Since our computational effort scales
as $d^{2 N_s}$, where $d$ is the Hilbert space dimensionality of the
subsystem, we are practically limited to $N_s <10$, implying on the
minimal time step that can be adopted.

\subsection{Markovian Master Equation}

The dynamics of the model (\ref{eq:H}), and its variants, can be
analyzed in the weak subsystem-bath coupling limit under the
Markovian approximation \cite{radiation,ET}. The probabilities $P_n$
to occupy the $|n\rangle$ state of the subsystem, the quantum
impurity, satisfy the Master equation
\bea \dot P_n= \sum_m P_m k_{m \rightarrow n}-P_n\sum_m
k_{n\rightarrow m}, \label{eq:popul} \eea
where the transition rate from the state $|m\rangle$ to $|n\rangle$ ($m\neq n$ and $m,n$=0,1 here)
is additive in the $L$ and $R$ reservoirs, $k_{n\rightarrow
m}=k^L_{n\rightarrow m}+k^R_{n\rightarrow m}$, due to the linear
form of the interaction \cite{MasterD,MasterWu}.

This type of Kinetic equation has been used in many recent works for
investigating energy, spin, and charge transfer in open quantum
systems \cite{MitraMaster,MasterWu}. Particularly, it has been
recently adopted for modeling radiative energy transfer between
metals \cite{radiation,Motonen}, and for studying charge and energy
transfer phenomenology in mesoscopic systems
\cite{BrandesRev,Brandes1,Teemu,Teemuheat} and single molecules
\cite{Koch1,*Koch2,*Koch3,*Koch4}. It is thus important to test the
suitability and accuracy of this common and well-accepted
approximate scheme against exact results.

In steady-state, taking Eq. (\ref{eq:curr})  as a starting point,
one can show that in the weak coupling limit and under the Markovian approximation
the energy current across the system reduces to \cite{Wu}
($\langle J_L\rangle =J$ in steady-state),
%
\bea
J=\sum_{m,n}E_{m,n}P_nk^L_{n\rightarrow m}, \label{eq:current}
\eea
with $E_{m,n}=E_m-E_n$. The current is defined positive when flowing
left to right. At the level of the Golden-Rule formula, the
transition rates are given by \cite{radiation}
\bea &&k_{n\rightarrow m}^\nu =
\nonumber\\
&& 2\pi \sum_{j,j'} |\alpha_{\nu,j;\nu,j'}|^2
 n_F^{\nu}(\epsilon_k) [1-n_F^{\nu}(\epsilon_{j'})]
\delta(\epsilon_{j}-\epsilon_{j'}-E_{m,n})
\nonumber\\
&&=2\pi  \int d \epsilon n_F^{\nu}(\epsilon)
[1-n_F^{\nu}(\epsilon-E_{m,n})] F_{\nu}(\epsilon).
\nonumber\\
&&=-2 \pi  n_B^{\nu}(E_{m,n}) \int d\epsilon \left[
n_F^{\nu}(\epsilon)-n_F^{\nu}(\epsilon-E_{m,n})
\right]F_{\nu}(\epsilon). \label{eq:FGR} \eea
From the last relation  we note that the thermal properties of the
reservoirs are concealed within both the Fermi-Dirac distribution
function
$n_F^{\nu}(\epsilon)=[e^{(\epsilon-\mu_{\nu})/T_{\nu}}+1]^{-1}$ and
the Bose-Einstein occupation factor
$n_B^{\nu}(\epsilon)=[e^{\epsilon/T_{\nu}}-1]^{-1}$. It is therefore
clear that when the integral yields a temperature independent
constant, the statistic of the reservoirs is fully bosonic
\cite{radiation}. The other element in Eq. (\ref{eq:FGR}) is a
dimensionless interaction term
\bea
F_{\nu}(\epsilon)=|\alpha_{\nu}|^2\rho_{\nu}(\epsilon)\rho_{\nu}(\epsilon-E_{m,n}),
\eea
which encloses the properties of the reservoirs, multiplied by the
subsystem-bath (energy independent) couplings $\alpha_{\nu}$.
Once we assume that the
density of states is a constant \cite{Persson},
$F_{\nu}(\epsilon)\approx F_{\nu}(\mu_{\nu})$, the integration in
Eq. (\ref{eq:FGR}) can be performed when the Fermi energies are situated far
from the conduction band edges
\cite{Persson}. Making use of the following relation,
\bea &&\int_{-\infty}^{\infty} d \epsilon
[n_F^{\nu}(\epsilon)-n_F^{\nu}(\epsilon-E_{m,n})] = -E_{m,n},
 \eea
we reach the closed-form expression,
\bea k^{\nu}_{n \rightarrow m}= 2 \pi n_{B}^{\nu}(E_{m,n})E_{m,n}
 F_{\nu}(\mu_{\nu}).
\label{eq:kPersson} \eea
Note that $n_{B}(-E_{m,n})=-[n_B(E_{m,n})+1]$, thus the excitation
and relaxation rates induced by the $\nu$ reservoir satisfy the
detailed balance relation, $k_{n\rightarrow m}^{\nu}/k_{m\rightarrow
n}^{\nu} = e^{-E_{m,n}/T_{\nu}}$. We can also express the rates in
terms of a subsystem-bath interaction parameter
\bea
\Gamma_F^{\nu}(2\Delta)&\equiv &2 \pi F_{\nu}(\mu_{\nu})2\Delta =
2\frac{2\Delta}{\pi}[\pi\rho_{\nu}(\mu_{\nu})\alpha_{\nu}]^2,
\label{eq:Gamma}
\eea
where we recall that both the density of states and the interaction
parameter $\alpha$ are assumed to be energy independent. Using this defining,
the rate constants in our model reduce to
\bea
k_{1\rightarrow 0}^{\nu} &=& \Gamma_F^{\nu}(2\Delta)
[1+n_B^{\nu}(2\Delta)],
\nonumber\\
k_{0\rightarrow 1}^{\nu}&=& \Gamma_F^{\nu}(2\Delta)
n_B^{\nu}(2\Delta). \label{eq:rate} \eea
For simplicity, we do not include below the explicit dependence of
$\Gamma_F^{\nu}$ and  $n_B^{\nu}$ on energy; both quantities should
be evaluated at the subsystem energy gap $2\Delta$. We calculate the
population of the states in steady-state by putting $\dot P_n=0$ in
Eq. (\ref{eq:popul}). With this at hand, the energy current
(\ref{eq:current}) simplifies to \cite{SegalR1,*SegalR2}
\bea J=2\Delta\frac{\Gamma_F^L\Gamma_F^R \left[n_B^L-n_B^R\right]}
{\Gamma_F^L(1+2n_B^L) +  \Gamma_F^R(1+2n_B^R) }. \label{eq:currM}
\eea
This expression provides the steady-state energy current in
the weak coupling limit, under the Markovian approximation.

It should be noted that a more involved non-interacting
blip-approximation (NIBA) type scheme \cite{SB} can be implemented
for following the qubit dynamics in the nonequilibrium spin-fermion
model \cite{MitraSpin1}. Furthermore, besides the qubit dynamics
itself, the heat current can be simulated within the NIBA
approximation by extending a generating function technique developed
in Ref. \cite{NIBASB} for the study of transport behavior in the
nonequilibrium spin-boson model. However, here we contain ourselves
with the simpler, more standard and common Markovian Master
equation, with the objective to provide insight on its applicability
and accuracy for the large community adopting it in studies of
charge, spin, and energy dynamics in mesoscopic and molecular
systems.

\begin{figure}[htbp]
{\hbox{\epsfxsize=80mm \hspace{0mm}\epsffile{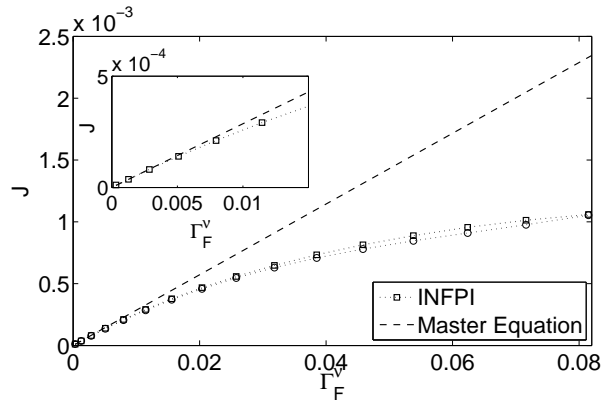}}}
\caption{Energy current as a function of metal-qubit coupling
parameter $\Gamma_F^{\nu}$ using the bandwidth $D=2$, $\Delta =0.1$,  $T_L=0.4$,
and $T_R=0.2$. INFPI numerical parameters are $\delta t=1$ and $N_s=9$.
Dashed line: Master equation results. INFPI results appear in symbols, $\square$
for $L=40$ and $\circ$ when taking the asymptotic $L\rightarrow \infty$ limit.
Inset: Zooming over the small coupling limit where the current linearly scales with $\Gamma_F^{\nu}$.
} \label{FigJ1}
\end{figure}

\begin{figure}[htbp]
{\hbox{\epsfxsize=80mm \hspace{0mm}\epsffile{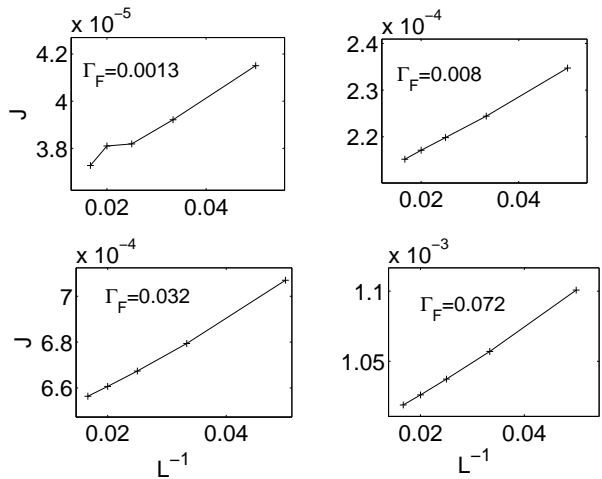}}}
\caption{
Converging the data of Fig. \ref{FigJ1} to the $L^{-1}\rightarrow 0$ limit,
with the intercept representing the asymptotic result.
} \label{FigN}
\end{figure}

\begin{figure}[htbp]
{\hbox{\epsfxsize=80mm \hspace{0mm}\epsffile{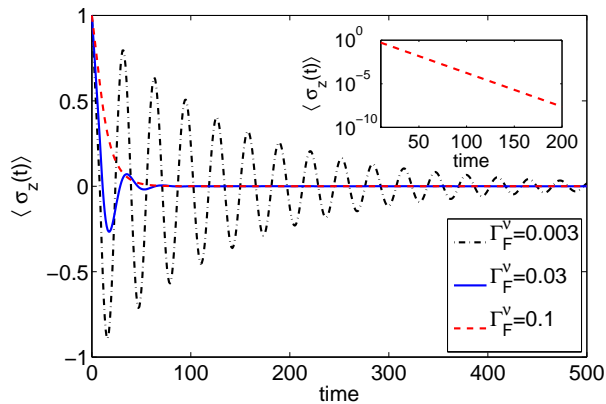}}}
\caption{Polarization dynamics at different coupling strengths,
for the same set of parameter
as in Fig. \ref{FigJ1}.
Results are displayed in the basis of the Hamiltonian
(\ref{eq:HH})
}
\label{FigTLS1}
\end{figure}

\section{Results}

We compare INFPI simulations to the Master equation predictions.
Within INFPI, the current is simulated directly using Eqs.
(\ref{eq:Ast}) and (\ref{eq:curr}), and we show results only in the
long time (quasi) steady-state limit. The closed-form Master
equation expression is given by (\ref{eq:currM}). Beyond the weak
coupling limit, we calculate the Master equation current directly
from Eq. (\ref{eq:current}) using the rates (\ref{eq:FGR}), since
the wide-band approximation does not hold anymore (though of course
the applicability of the Master equation technique as a whole is
questionable in this regime).

\begin{figure}[htbp]
{\hbox{\epsfxsize=80mm \hspace{0mm}\epsffile{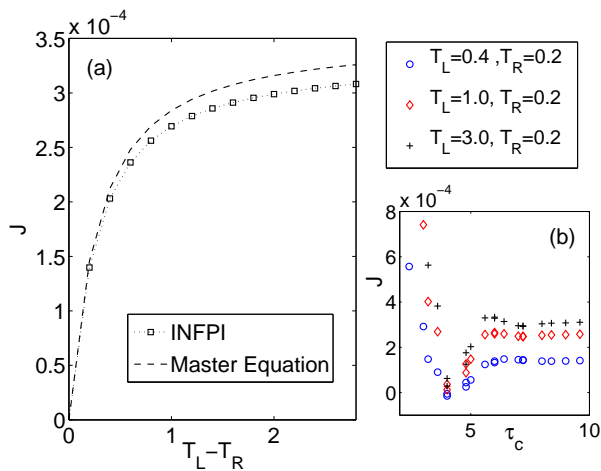}}}
\caption{(a) Energy current-temperature characteristics
for the same set of parameter as in Fig. \ref{FigJ1},
$\phi=0.2$ ($\Gamma_F=0.005$). We vary $T_L$, but keep $T_R$ fixed, $T_R=0.2$.
(b) Convergence behavior with increasing memory size.
Data was produced with three different time steps,
$\delta t=$ 0.8, 1.0 and 1.2.
}
\label{FigT1}
\end{figure}

\begin{figure}[htbp]
{\hbox{\epsfxsize=80mm \hspace{0mm}\epsffile{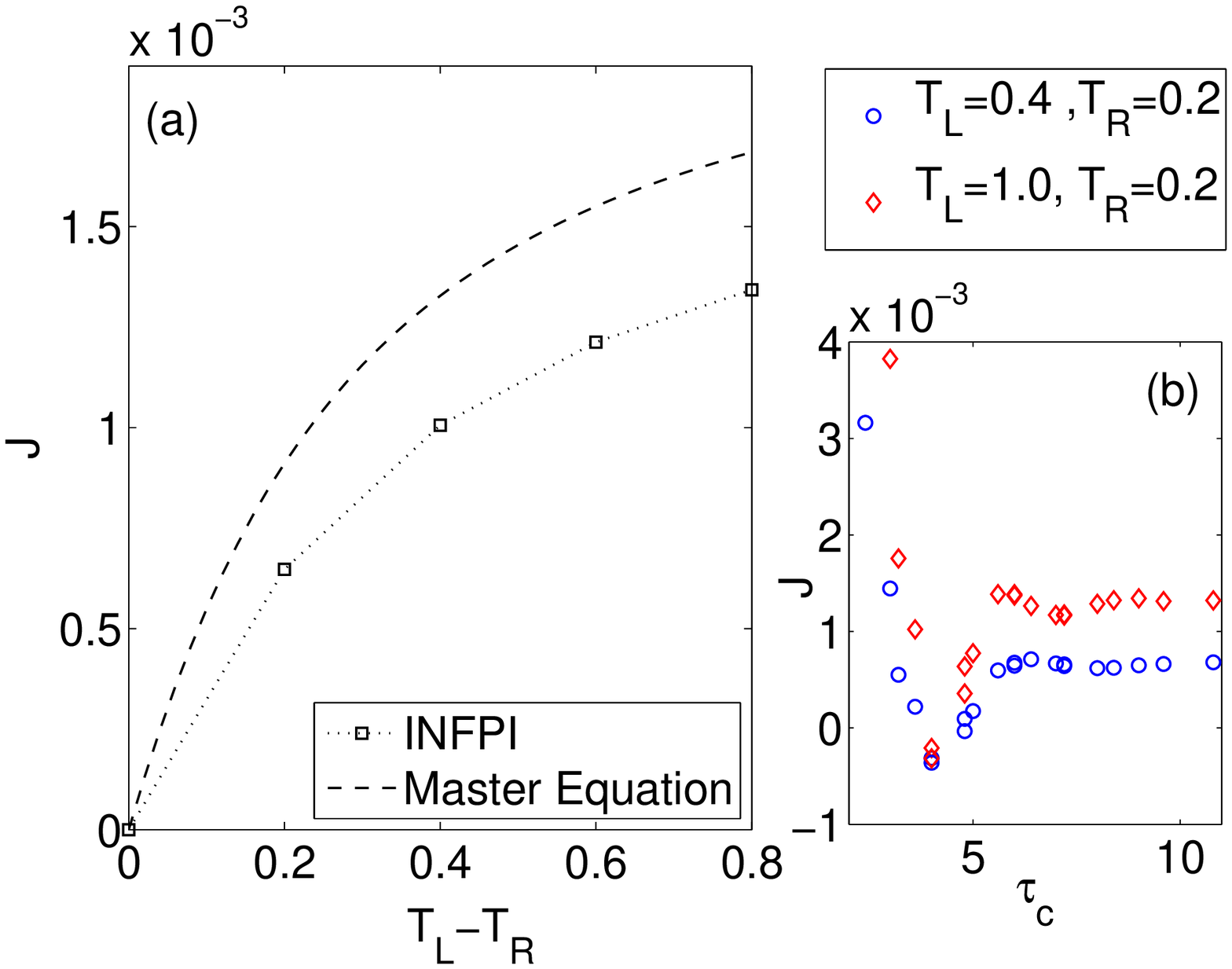}}}
\caption{(a) Energy current-temperature characteristics
for the same set of parameter as in Fig. \ref{FigJ1},
$\phi=0.5$ ($\Gamma_F=0.032$). We vary $T_L$, but keep $T_R$ fixed, $T_R=0.2$.
(b) Convergence behavior with increasing memory size.
Data was produced with three different time steps,
$\delta t=$ 0.8, 1.0 and 1.2.
}
\label{FigT2}
\end{figure}

\begin{figure}[htbp]
{\hbox{\epsfxsize=80mm \hspace{0mm}\epsffile{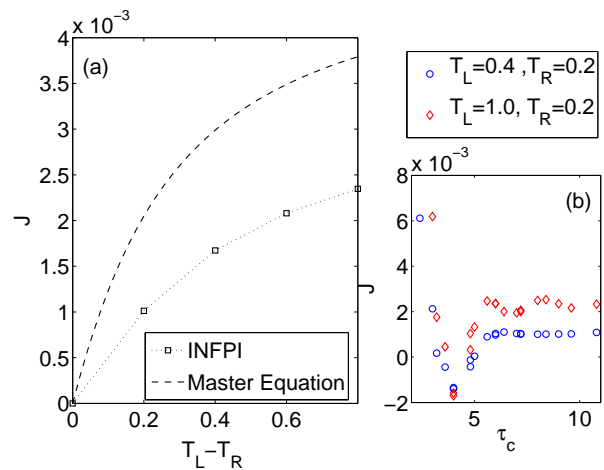}}}
\caption{(a) Energy current-temperature characteristics
for the same set of parameter as in Fig. \ref{FigJ1},
$\phi=0.75$ ($\Gamma_F=0.072$). We vary $T_L$, but keep $T_R$ fixed, $T_R=0.2$.
(b) Convergence behavior with increasing memory size:
results converge only at small bias, $T_L-T_R<0.4$.
Data was produced with three different time steps,
$\delta t=$ 0.8, 1.0 and 1.2.
}
\label{FigT3}
\end{figure}

We typically use the following set of parameters: an energy gap
$2\Delta=0.2$ (arbitrary units) for the subsystem, metallic
bandwidth $D=2$, $T_{\nu}\gtrsim\Delta$, and the equilibrium Fermi
energies located at the
center of the band. 
We also define the dimensionless parameter
\bea
\phi_{\nu}=\pi\rho_{\nu}\alpha_{\nu},
\eea
which is varied between 0 and 0.8 here, where convergence is
achieved. This choice corresponds to $0<\Gamma_F^{\nu}<0.08$ when
$\Delta=0.1$, see Eq. (\ref{eq:Gamma}). For this set of model
parameters, we have confirmed that selecting
$\delta t=0.8-1.5$ and $N_s= 6-9$ (yielding memory time
$\tau_c \geq 8$) provides converging results, see panel b in Figs. \ref{FigT1}, \ref{FigT2}
and \ref{FigT3}.

Before turning to simulations, it is important to identify the
region of weak subsystem-bath coupling.
It holds 
when the scattering phase shift is small, $\arctan \phi_{\nu}\sim
\phi_{\nu}$ \cite{Nozieres1,*Nozieres3}. This translates (within
$5\%$ error) to the dimensionless parameter $\phi$ being limited to
the domain of  $\phi_{\nu} \lesssim 0.4$. Equivalently, the
interaction energy  should be limited to $\Gamma_F^{\nu}\lesssim
0.02$, or in other words, $\Gamma_F^{\nu}/(2\Delta)\lesssim0.1$,
i.e., in the weak coupling limit the subsystem gap is large relative
to the coupling energy.


Fig. \ref{FigJ1} displays the energy current as a function of the
interaction energy for a symmetric junction,
$\Gamma_F^{L}=\Gamma_F^{R}$. We find that when
$\Gamma_F^{\nu}\gtrsim 0.02$, Master equation-derived energy current
overestimates the exact result by more than 10$\%$. In the strong
coupling limit, Master equation overvalues the correct numbers by a
factor of two. More significantly, this Golden-Rule based method
cannot reproduce the saturation effect of the current with the
subsystem-bath interaction parameter, and it wrongly predicts a
linear scaling, $J\propto\Gamma_F^{\nu}$. Note that we include INFPI
results both for the case of $L=40$ bath states ($\square$), and in
the asymptotic $L\rightarrow \infty$ limit ($\circ$), obtained by
extrapolating the linear $J$ vs. $L^{-1}$ curves to
$L^{-1}\rightarrow 0$, see Fig. \ref{FigN}. We find that this
extrapolation affects the results by up to $4\%$ at strong coupling,
while the weak coupling values are unaffected. While we do not show
 transient data for the current, we comment that steady-state has been reached at
$\Delta t\sim 50-100$ in the weak coupling limit; it is established
much faster, $\Delta t\sim 5-10$ at strong coupling.

We correlate transport behavior of the junction with a study of the
dissipative dynamics of the spin polarization, as obtained from
INFPI, in Fig. \ref{FigTLS1}. We observe weakly-damped coherent
oscillations in the weak coupling limit when perturbative Master
equation well describes the dynamics. These oscillations still
survive at intermediate coupling, but at strong coupling the
polarization exponentially decays in time (inset). Crucial
parameters of the model are the scattering phase shifts
$\delta_{\pm}$. In equilibrium, the phase shifts are given by
\cite{Nozieres1,*Nozieres3,Ng}
\bea \tan \delta_{\pm}= \phi_{L,R}
\label{eq:phaseE} \eea
Out-of-equilibrium, $\Delta\mu \neq 0$, the phase shifts are {\it
complex} numbers \cite{Ng}. In the spin-boson model the Kondo
dimensionless dissipation coefficient $\xi$ represents a
characteristic exponent in the system: at zero temperature and zero
energy bias the spin displays damped coherent oscillations for
$\xi<0.5$, relaxation dynamics between $0.5\leq\xi<1$, and a
localization phase for $\xi\geq1$ \cite{LeHur}. Thus, this parameter
controls dissipation-induced phase transitions. In the fermionic
analogue it can be shown that the characteristic exponent is given
by $\xi=(\delta_+^2+\delta_-^2)/\pi^2$ \cite{Ng}. Since
$|\delta_{\pm}|\leq \pi/2$, $\xi\leq1/2$. Thus, in the spin-fermion
model described in this paper the spin cannot manifest the
localization behavior, and a large $\phi$ value brings us to the
relaxation scenario, as indeed observed in Fig. \ref{FigTLS1}.



The current-temperature characteristics of the junction are depicted
in Figs. \ref{FigT1}-\ref{FigT3}, using different coupling
strengths. We find that at weak coupling Markovian Master equation
very well reproduces the qualitative and quantitative aspects of the
current, even at a large temperature difference, see Fig.
\ref{FigT1}. The convergence behavior of the INFPI method at
different temperature biases is displayed in panel (b), where we
show the energy current as obtained using different memory time
$\tau_c$ and time steps. At intermediate couplings, Fig. \ref{FigT2}
shows that Master equation overestimates exact results by up to
25$\%$ at large temperature differences. When the subsystem-bath
coupling is large, we managed to converge simulations only up to the
bias $T_L-T_R\sim 0.2$. The Kinetic method now provides values that
are a factor of 2 larger than the exact numerical data. It is
important to note that the qualitative current-temperature features
are correctly reproduced within the Markovian Master equation, even
at strong coupling. However, if one is interested in quantitative
information, Master equation can be used in its strict regime of
applicability only, $\phi\lesssim 0.2$. Data in Figs.
\ref{FigT1}-\ref{FigT3} is presented for a fixed value of bath
states, $L=40$, since we have confirmed that taking the large-$L$
limit only corrects the current by $\lesssim 4\%$, at both low and
high temperature biases.

\section{Summary}

We have studied energy transfer between metals mediated by a quantum
impurity, using two approaches: numerically exact path-integral
simulations and analytic results from a Golden-Rule type Markovian
Master equation treatment. We found that standard Master equations
fail to reproduce the current-interaction energy characteristics
already at intermediate system-bath couplings, as it can only
provide a linear enhancement of the current with the subsystem-bath
interaction, missing a saturation effect. In contrast, the
current-temperature characteristics is produced in a qualitative
correct way by a Master equation formalism, though actual values
deviate by 100$\%$, and more,  at high temperature biases and at
strong coupling.

Our results are beneficial for the critical testing of common Master
equation techniques. The methods described are also useful for
practically modeling superconducting-based qubit devices
\cite{PekolaRev}. While a Master equation treatment offers
simple-intuitive expressions that often allow to discern essential
transport characteristics, already at intermediate system-bath
couplings it may overestimate the current by $\sim 10 \%$, up to a
$100\%$ incorrect enhancement at strong coupling. These deviations
are certainly important when a quantitative analysis of device
efficiency is performed. In particular, the calculation of energy
conversion efficiency in conducting junctions should be done with
caution when a Master equation is of use
\cite{Brandes1,OjanenEff,esposito1}.

In our future work we plan to study the heat current characteristics in the complementary
spin-boson type molecular junction model \cite{SegalR1,*SegalR2}. This could be done
 by extending the Feynman-Vernon IF expression \cite{FV} to describe the evolution
of other operators besides the reduced density matrix. Alternatively, one could use the
bosonization approach and draw general results for the nonequilibrium
spin-boson problem, based on the spin-fermion model calculations presented here.

\begin{acknowledgments}
Support from an NSERC discovery grant is acknowledged.
\end{acknowledgments}

\bibliography{SF2}
\end{document}